\documentclass[12pt]{article}
\let\tilde=\widetilde

\def\phi{\varphi}

\def\d{\delta}

\def\l{\lambda}
\def\L{\Lambda}

\def\tr{\mathop{\hbox{\rm tr}}\nolimits}

\def\res{\mathop{\hbox{\rm res}}\limits}

\def\dd{\partial}

\def\one#1{#1^{\raise5pt\hbox{$\scriptstyle\!\!\!\!1$}}\,{}}
\def\two#1{#1^{\raise5pt\hbox{$\scriptstyle\!\!\!\!2$}}\,{}}
\def\three#1{#1^{\raise5pt\hbox{$\scriptstyle\!\!\!\!3$}}\,{}}
\def\ket#1{\left|#1\right>}
\def\bra#1{\left<#1\right|}

\def\id{\hbox{{1}\kern-.25em\hbox{\rm l}}}

\def\comment#1{}

\def\beq{\begin{equation}}
\def\eeq{\end{equation}}
\def\be{\begin{displaymath}}
\def\ee{\end{displaymath}}
\def\bea{\begin{eqnarray}}
\def\eea{\end{eqnarray}}
\def\beas{\begin{eqnarray*}}
\def\eeas{\end{eqnarray*}}
\def\bds{\begin{description}}
\def\eds{\end{description}}
\def\bmat{\left(\begin{array}}
\def\emat{\end{array}\right)}

\def\Ref#1{(\ref{#1})}
\def\?{(?)\marginpar{|?}}

\newcounter{subequation}[equation]
\makeatletter
\expandafter\let\expandafter
\reset@font\csname reset@font\endcsname
\def\subeqnarray{\arraycolsep1pt
    \def\@eqnnum\stepcounter##1{\stepcounter{subequation}%
        {\reset@font\rm(\theequation\alph{subequation})}}
\jot5mm     \eqnarray}

\makeatother

\newcommand\ftnote[1]{\setcounter{footnote}{#1}\addtocounter{footnote}{-1}
\footnote}
\newfont{\bbd}{msbm10 scaled\magstep1} 
\def\C{\hbox{\bbd C}}                  

\def\P{{\cal P}}
\def\id{\hbox{1\hskip-3pt{\rm l}}}
\def\bt{B\"acklund transformation}
\def\Sb{\hbox{\bbd S}}
\begin{document}
\begin{flushright}
\sf LPENSL-Th 05/99 \\
\sf solv-int/9903016 \\
\end{flushright}
\vskip1cm
\begin{center}\LARGE\bf
Canonicity of \bt:
$r$-matrix approach. I.
\end{center}
\vskip1cm
\begin{center}
E K Sklyanin\ftnote{1}{On leave from: Steklov Mathematical Institute at
St.~Petersburg, Fontanka 27, St.~Petersburg 191011, Russia. 
E-mail: {\tt sklyanin\symbol{'100}euclid.pdmi.ras.ru}}
\vskip0.4cm
Laboratoire de Physique\ftnote{2}{UMR 5672 du CNRS et de l'ENS Lyon}, 
Groupe de Physique Th\'eorique, ENS Lyon, \\
46 all\'ee d'Italie, 69364 Lyon 07, France
\end{center}
\vskip2cm
{\bf Abstract.} For the Hamiltonian integrable systems governed by 
$SL(2)$-invariant $r$-matrix (such as Heisenberg magnet, Toda lattice,
nonlinear Schr\"{o}dinger equation)  a general procedure for constructing 
\bt\ is proposed. The corresponding BT is shown to preserve the  Poisson
bracket. The proof is given by a direct calculation using the $r$-matrix 
expression for the Poisson bracket.
\vskip1cm
\begin{center}
 25 March, 1999
\end{center}

\vskip2cm

\newpage


In 1976 Flaschka and McLaughlin \cite{FM76} has demonstrated that the 
standard \bt\ (BT) for the KdV equation is canonical with respect to the 
associated symplectic structure. Subsequently, the canonicity of BT{s}
has been proved for some more integrable models, see e.\ g.\ Appendix to
\cite{MS91}.

The aim of the present paper is to apply the $r$-matrix formalism \cite{FT87}
to the problem of proving the canonicity of BT. Our proof is quite 
general and requires only that the Poisson brackets of the corresponding
Lax operator could be written down in the $r$-matrix form. In our treatment
of \bt\ we adopt the approach of \cite{KS5} where a program has been formulated
of reexamining BT{s} from the Hamiltonian point of view.

Consider an integrable Hamiltonian system possessing a Lax matrix $L(u)$
depending on the dynamical variables and a complex parameter $u$ (spectral
parameter). The spectral invariants of $L(u)$ are supposed  to generate the
commuting Hamiltonians of the system. Defining a \bt\ as  a canonical
transformation preserving the Hamiltonians of the system (see \cite{KS5} for
a more detailed list of properties of BT) we conclude that it has  to preserve
as well the spectral invariants of $L(u)$. As a consequence, the original
matrix $L(u)$ and the transformed one $\L(u)$ must be related by a
similarity transformation
\beq
 \L(u)=M(u)L(u)M(u)^{-1}.
\label{def-Lambda}
\eeq
(see \cite{MS91} for a detailed account of the theory of BT as
gauge transformations).

An important practical question arising in the theory of BT is how to find, 
given a Lax matrix $L(u)$, such a matrix $M(u)$ which would generate a BT.
To check that a matrix $M(u)$ is admissible one needs, first, to verify that
the system of equations resulting from \Ref{def-Lambda} is self-consistent,
and, second, to proof that the resulting transformation of dynamical
variables is canonical, that is preserves the Poisson bracket.

Below we solve the both problems for the class of integrable systems governed 
by the $SL(2)$-invariant $r$-matrix. Suppose that $L(u)$ is a matrix of order
$2\times2$ and the Poisson brackets between the entries of $L(u)$ can be
expressed in the $r$-matrix form \cite{FT87}
\beq
 \{\one L(u),\two L(v)\}=[r_{12},\one L(u)\two L(v)]
\label{pb-LL}
\eeq
where
$\one L=L\otimes\id$, $\two L=\id\otimes L$, and
$ r_{12}=\kappa(u-v)^{-1}\P_{12}$
is the standard $SL(2)$-invariant solution to the classical Yang-Baxter 
equation \cite{FT87}, $\kappa$ being a constant and $\P_{12}$ being the
permutation operator in $\C^2\otimes\C^2$.
The class of integrable models thus defined includes such well-known models
as the nonlinear Schr\"odinger equation, Heisenberg magnetic chain, Toda 
lattice \cite{FT87}.

When choosing an ansatz for the matrix $M(u)$ we shall follow 
\cite{QNCL84,NP89} where it was observed that, in the cases of Heisenberg
magnetic chain and of the lattice Landau-Lifshitz equation, the matrix $M(u)$ 
happens to have the same form, as a function of $u$, as the corresponding
elementary Lax matrix $L(u)$ for the chain consisting only of one atom. 
As shown below, such choice of $M(u)$ is valid for any integrable model
governed by the $r$-matrix specified above.

Our ansatz  for $M(u)$ mimics the form of elementary Lax matrix for the
isotropic Heisenberg magnetic chain \cite{FT87,QNCL84}:
\beq
 M(u)=(u-\l)\id+\Sb
\label{eq:ansatz-M}
\eeq
where
\beq
  \tr\Sb=0, \qquad \det\Sb=-\mu^2
\label{eqs-S}
\eeq
$\l$ and $\mu$ being free parameters. It is convenient to perform a 
reparametrization  $\l_1=\l+\mu$, $\l_2=\l-\mu$, so that $\mu=(\l_1-\l_2)/2$.
The constraints \Ref{eqs-S}  on the matrix $\Sb$ leave two more undetermined
parameters. Denoting them $p$ and $q$ and choosing a particular parametrization
of $\Sb$ we fix the following ansatz for $M(u)$
\beq
   M(u)=\left(\begin{array}{cc}
         u-\l_1+pq & p \\ -pq^2+2\mu q & u-\l_2-pq
         \end{array}\right).
\label{def-M}
\eeq

We shall consider $\l_1$, $\l_2$ as the free parameters of BT. The parameters
$p$, $q$ are then to be determined from the equations \Ref{def-Lambda}.

Let us introduce the eigenbasis of $M(u)$
\beq
 \ket{1}=\frac{1}{2\mu}\left(\begin{array}{c} 1 \\ -q  \end{array}\right),
 \qquad
   \ket{2}=\frac{1}{2\mu}\left(\begin{array}{c} p \\ 2\mu-pq
                          \end{array}\right)
\eeq
and the dual eigenbasis
\beq
 \bra{1}=(2\mu-pq,\,-p), \qquad
 \bra{2}=(q,\,1),
\eeq
as well as the corresponding spectral projectors
\beq
    P_{ij}= \ket{i}\bra{j}, \qquad i,j\in\{1,2\}
\eeq
satisfying
\beq
    P_{ij}P_{kl}=P_{il}\d_{jk}.
\eeq

In terms of the projectors $P_{ij}$ the matrix $M(u)$  and its inverse
are written down, respectively,  as
\beq
  M(u)=(u-\l_1)P_{11}+(u-\l_2)P_{22}
\eeq
and
\beq
  M(u)^{-1}=(u-\l_1)^{-1}P_{11}+(u-\l_2)^{-1}P_{22}.
\label{M^{-1}}
\eeq

Note that
\beq
    \det M(u)=(u-\l_1)(u-\l_2).
\eeq

Let us derive now from \Ref{def-Lambda} the equations determining the
parameters $p$, $q$. Suppose that $L(u)$ is polynomial in $u$ (this covers all
lattice models, the continuous models can then be obtained as appropriate 
limits). Requiring that the transformation \Ref{def-Lambda} preserves
the polynomiality of $L(u)$ we conclude that the apparent poles  of
the right-hand-side of \Ref{def-Lambda} at $u=\l_{1,2}$ due to
\Ref{M^{-1}} should vanish,
\beq
 0=\res_{u=\l_i}\L(u)=(\l_i-\l_{1-i})P_{1-i,1-i}L(\l_i)P_{ii},
\eeq
which gives us two 
equations for determining $p$ and $q$ as
functions of the dynamical variables of the system:
\beq
 \tr P_{12}L(\l_1)=0, \qquad \tr P_{21}L(\l_2)=0.
\label{eq-pq}
\eeq

With the parameters $p$ and $q$ determined by the equations \Ref{eq-pq},
the matrix $\L(u)$ is defined by \Ref{def-Lambda} as a function of the
dynamical variables and free parameters $\l_{1,2}$.
The next step is to show that
$\L(u)$ satisfies the same Poisson bracket relations \Ref{pb-LL} as $L(u)$:
\beq
 \{\one\L(u),\two\L(v)\}=[r_{12},\one\L(u)\two\L(v)].
\label{pb-LmbLmb}
\eeq

The calculation of the Poisson brackets for $\L(u)$, though cumbersome,
is quite straightforward, since all the necessary ingredients are already 
prepared. Substituting \Ref{def-Lambda} into $\{\one\L(u),\two\L(v)\}$ and
differentiating the products of matrices we obtain a rather long expression.
To write it down in a more compact form let us introduce the following 
notation
\begin{subeqnarray}
 \left<\one L\two M\right>&=&
 \one M(u)\{\one L(u),\two M(v)\}\one M(u)^{-1}\two M(v)^{-1},\\
 \left<\one M\two L\right>&=&
 \two M(v)\{\one M(u),\two L(v)\}\one M(u)^{-1}\two M(v)^{-1},\\
   \left<\one M\two M\right>&=&
 \one M(u)\two M(v)\{\one M(u),\two M(v)\}\one M(u)^{-1}\two M(v)^{-1},
\label{def<>}
\end{subeqnarray}
\beq
  \tilde r_{12}=\one M(u)\two M(v)r_{12}\one M(u)^{-1}\two M(v)^{-1}.
\label{def-rtilde}
\eeq

Using \Ref{def<>}, \Ref{def-rtilde} one can write down the left-hand-side
of \Ref{pb-LmbLmb} as
\bea
 \{\one\L(u),\two\L(v)\}&=&
 \left<\one M\two M\right>\one\L(u)\two\L(v)
+\left<\one L\two M\right>\two\L(v)
-\one\L(u)\left<\one M\two M\right>\two\L(v) \nonumber \\
&&+\left<\one M\two L\right>\one\L(u)
+[\tilde r_{12},\one\L(u)\two\L(v)]
-\one\L(u)\left<\one M\two L\right> \nonumber \\
&&-\two\L(v)\left<\one M\two M\right>\one\L(u)
-\two\L(v)\left<\one L\two M\right>
+\one\L(u)\two\L(v)\left<\one M\two M\right>.
\label{9terms}
\eea

Our aim is to show that the resulting expression is equal to 
$[r_{12},\one\L(u)\two\L(v)]$. Note, first, that using the identity
\be
 \P_{12}=\one P_{11}\two P_{11}+\one P_{12}\two P_{21}+\one P_{21}\two P_{12}+
          \one P_{22}\two P_{22}
\ee
one can show that
\beq
  \tilde r_{12}=r_{12}+2\mu\kappa
\left(\frac{\one P_{12}\two P_{21}}{(u-\l_2)(v-\l_1)}
 -\frac{\one P_{21}\two P_{12}}{(u-\l_1)(v-\l_2)}\right).
\label{r-tilde}
\eeq

It remains then to calculate the Poisson brackets between $L(u)$ and $M(u)$.
To do it, we recollect the equations \Ref{eq-pq} defining implicitely $p$
and $q$. The calculate the Poisson brackets for $p$ and $q$ we shall
use the trick employed in a similar situation in \cite{Skl38}.
For any function $f$ on  the phase  space we have
\begin{subeqnarray}
0&=&\{f,\tr P_{12}L(\l_1)\} \nonumber \\
&=& \{f,p\}\tr\frac{\dd P_{12}}{\dd p}L(\l_1)
+\{f,q\}\tr\frac{\dd P_{12}}{\dd q}L(\l_1)
+\tr P_{12}\{f,L(\l_1)\},\\
0&=&\{f,\tr P_{21}L(\l_2)\} \nonumber \\
&=& \{f,p\}\tr\frac{\dd P_{21}}{\dd p}L(\l_2)
+\{f,q\}\tr\frac{\dd P_{21}}{\dd q}L(\l_2)
+\tr P_{21}\{f,L(\l_2)\}.
\label{pb-fpq}
\end{subeqnarray}

The Poisson brackets $\{f,p\}$ and $\{f,q\}$ are then determined by solving the
linear system \Ref{pb-fpq}. Using the equalities
\beq
 \frac{\dd P_{12}}{\dd p}=0, \qquad
 \frac{\dd P_{12}}{\dd q}=\frac{1}{2\mu}(P_{11}-P_{22})+\frac{p}{\mu}P_{12},
\eeq
\beq
 \frac{\dd P_{21}}{\dd p}=P_{11}-P_{22}, \qquad
 \frac{\dd P_{21}}{\dd q}=\frac{p^2}{2\mu}(P_{11}-P_{22})-\frac{p}{\mu}P_{21},
\eeq
\beq
  \frac{\dd M(u)}{\dd p}=2\mu p, \qquad
  \frac{\dd M(u)}{\dd q}=p^2P_{12}+P_{21}
\eeq
and introducing the notation
\beq
 w_i(\l)=\tr P_{ii}L(\l), \qquad w(\l)=w_1(\l)-w_2(\l)
\eeq
one obtains then
\beq
 \{f,M(v)\}=-\frac{2\mu}{w(\l_1)}P_{21}\tr\{f,L(\l_1)\}P_{12}
            -\frac{2\mu}{w(\l_2)}P_{12}\tr\{f,L(\l_2)\}P_{21}
\eeq
(note that the last formula does not depend on parametrization of \Sb).

Now, using \Ref{pb-LL}  it is easy to calculate the brackets
\bea
\{\one L(u),\two M(v)\}&=&
-\frac{2\mu\kappa}{(u-\l_1)w(\l_1)}
\left(w_1(\l_1)\one P_{12}\one L(u)\two P_{21}
     -w_2(\l_1)\one L(u)\one P_{12}\two P_{21}\right)\nonumber \\
&&-\frac{2\mu\kappa}{(u-\l_2)w(\l_2)}
\left(w_2(\l_2)\one P_{21}\one L(u)\two P_{12}
     -w_1(\l_2)\one L(u)\one P_{21}\two P_{12}\right),
\eea

\bea
\{\one M(u),\two L(v)\}&=&
\frac{2\mu\kappa}{(v-\l_1)w(\l_1)}
\left(w_1(\l_1)\one P_{21}\two P_{12}\two L(v)
     -w_2(\l_1)\one P_{21}\two L(v)\two P_{12}\right)\nonumber \\
&+&\frac{2\mu\kappa}{(v-\l_2)w(\l_2)}
\left(w_2(\l_2)\one P_{12}\two P_{21}\two L(v)
     -w_1(\l_2)\one P_{12}\two L(v)\two P_{21}\right),
\eea

\beq
 \{\one M(u),\two M(v)\}=
 -\frac{2\mu\kappa\bigl(w_1(\l_1)w_2(\l_2)-w_2(\l_1)w_1(\l_2)\bigr)}%
{w(\l_1)w(\l_2)}
 \left(\one P_{12}\two P_{21}-\one P_{21}\two P_{12}\right),
\eeq

Recalling the notation \Ref{def<>} one arrives to the expressions
\bea
 \left<\one L\two M\right>&=&
-\frac{2\mu\kappa}{(u-\l_2)(v-\l_1)}\left(
 \frac{w_1(\l_1)}{w(\l_1)}\one P_{12}\one\L(u)\two P_{21}
-\frac{w_2(\l_1)}{w(\l_1)}\one\L(u)\one P_{12}\two P_{21}\right)\nonumber\\
&&-\frac{2\mu\kappa}{(u-\l_1)(v-\l_2)}\left(
 \frac{w_2(\l_2)}{w(\l_2)}\one P_{21}\one\L(u)\two P_{12}
-\frac{w_1(\l_2)}{w(\l_2)}\one\L(u)\one P_{21}\two P_{12}\right),
\label{LM}
\eea

\bea
  \left<\one M\two L\right>&=&
 \frac{\mu\kappa}{(u-\l_1)(v-\l_2)}\left(
 \frac{w_1(\l_1)}{w(\l_1)}\one P_{21}\two P_{12}\two\L(v)
-\frac{w_2(\l_1)}{w(\l_1)}\one P_{21}\two\L(v)\two P_{12}\right)\nonumber\\
&&+\frac{\mu\kappa}{(u-\l_2)(v-\l_1)}\left(
 \frac{w_2(\l_2)}{w(\l_2)}\one P_{12}\two P_{21}\two\L(v)
-\frac{w_1(\l_2)}{w(\l_2)}\one P_{12}\two\L(v)\two P_{21}\right),
\label{ML}
\eea

\bea
 \left<\one M\two M\right>&=&
-\frac{2\mu\kappa\bigl(w_1(\l_1)w_2(\l_2)-w_2(\l_1)w_1(\l_2)\bigr)}%
{w(\l_1)w(\l_2)}\nonumber\\
&&\times\left(\frac{\one P_{12}\two P_{21}}{(u-\l_2)(v-\l_1)}
     -\frac{\one P_{21}\two P_{12}}{(u-\l_1)(v-\l_2)}\right).
\label{MM}
\eea

Finally, substituting \Ref{LM}, \Ref{ML}, \Ref{MM} together with
\Ref{r-tilde} into \Ref{9terms}, we obtain, after massive cancellations,
the equality \Ref{pb-LmbLmb} which finishes the proof of the canonicity 
of \bt.

\section*{Discussion}

In this paper, we have studied only the case of the XXX type $r$-matrix.
There is little doubt that our results can be generalized to the cases of
XXZ and XYZ type $r$-matrices. Taking the limit of linear Poisson brackets
\beq
  \{\one{L}(u),\two{L}(v)\}=[r_{12},\one{L}(u)+\two{L}(v)]
\eeq
one obtains, as a corollary, the canonicity of BT for the Gaudin model
\cite{Skl38}.

The proof of canonicity of BT given in the present paper, being quite simple
and straightforward, does not explain, however, the reasons why we should
choose for $M$ the same ansatz \Ref{eq:ansatz-M} as for $L$.  An answer to
this question is given in the second part of our paper which is being prepared
for publication. 

\section*{Acknowledgements}

I am grateful to V.~B.~Kuznetsov and F.~Nijhoff for
their interest in the work and useful discussions. This work was started
during my stay at the Department of Mathematics, the University of Leeds.
and benefited from the financial support of EPSRC.


\end{document}